\documentclass[]{spie}  

\usepackage{amsmath,amsfonts,amssymb}
\usepackage{graphicx}
\usepackage[colorlinks=true, allcolors=blue]{hyperref}

\title{Joint optimization of wavefront sensing and reconstruction with automatic differentiation}

\author[a]{Rico Landman}
\author[a,b]{Christoph Keller}
\author[c]{Emiel H. Por}
\author[d]{Sebastiaan Haffert}
\author[a,e]{David Doelman}
\author[a]{Thijs Stockmans}
\affil[a]{Leiden Observatory, Leiden University, PO Box 9513, 2300 RA Leiden, The Netherlands}
\affil[b]{Lowell Observatory, 1400 W Mars Hill Rd, Flagstaff, AZ 86001, USA}
\affil[c]{Space Telescope Science Institute, Baltimore, MD 21218, USA}
\affil[d]{University of Arizona, Steward Observatory, Tucson, Arizona, United States}
\affil[e]{SRON Netherlands Institute for Space Research, Niels Bohrweg 4, 2333 CA Leiden, the Netherlandss}

\authorinfo{Further author information, send correspondence to R. Landman (rlandman@strw.leidenuniv.nl)}

\begin{document} 
\maketitle

\begin{abstract}
High-contrast imaging instruments need extreme wavefront control to directly image exoplanets. This requires highly sensitive wavefront sensors which optimally make use of the available photons to sense the wavefront. Here, we propose to numerically optimize Fourier-filtering wavefront sensors using automatic differentiation. First, we optimize the sensitivity of the wavefront sensor for different apertures and wavefront distributions. We find sensors that are more sensitive than currently used sensors and close to the theoretical limit, under the assumption of monochromatic light. Subsequently, we directly minimize the residual wavefront error by jointly optimizing the sensing and reconstruction. This is done by connecting differentiable models of the wavefront sensor and reconstructor and alternatingly improving them using a gradient-based optimizer. We also allow for nonlinearities in the wavefront reconstruction using Convolutional Neural Networks, which extends the design space of the wavefront sensor. Our results show that optimization can lead to wavefront sensors that have improved performance over currently used wavefront sensors. The proposed approach is flexible, and can in principle be used for any wavefront sensor architecture with free design parameters.
\end{abstract}

\keywords{Adaptive optics, wavefront sensing, high contrast imaging, machine learning, optimization}

\section{INTRODUCTION}
High contrast imaging systems require extreme wavefront precision to directly image extrasolar planets. To achieve this, wavefront aberrations due to the Earth's atmosphere or instrument optics need to be corrected through an Adaptive Optics (AO) system. This AO system uses a wavefront sensor (WFS) to sense the incoming aberrations and subsequently corrects this through a deformable mirror (DM). While early AO systems used Shack-Hartmann Wavefront Sensors (SHWFS), the current preference for ground-based high-contrast imaging is the Pyramid Wavefront Sensor (PWFS) \cite{1996JMOp...43..289R_ragazzoni_pwfs}. The PWFS belongs to the group of Fourier-filtering wavefront sensors (FFWFS) \cite{2016Optic...3.1440F_fauvarque_fourier_formalism, 2017JATIS_fauvarque_formalism_pwfs}. These FFWFS's use a phase and/or amplitude mask in the focal plane that filters the electric field and transforms phase fluctuations into intensity variations in the pupil plane. Another prominent example of a FFWFS is the Zernike Wavefront Sensor (ZWFS) \cite{2013A&A...555A..94N_ndiaye_zelda}. Recent work has studied alterations of the PWFS \cite{2015OptL...40.3528F_fauvarque_flattened_pwfs, 2021SPIE11823E..1BG_gerard_bright_pwfs,2017JATIS_fauvarque_formalism_pwfs, 2021JATIS...7d9001S_schatz_3pwfs} and ZWFS \cite{2021A&A...650L...8C_chambouleyron_zernike2}, and showed that these lead to improved sensitivity or performance under certain conditions. These alterations were chosen based on intuition, and were not the result of a global optimization procedure, such as commonly done for the design of coronagraphs \cite{2011ApJ...729..144S_soummer_coronagraph_optimization, 2013A&A...551A..10C_carlotti_coronagraph_optimzation, 2016ApJ...818..163N_ndiaye_coronagraph_optimization, 2017SPIE10400E_por_coronagraph_optimization}. This is a complicated trade-off between, among others, the sensitivity, dynamic range and photon efficiency for different spatial frequencies. Furthermore, these WFS are often designed to have an as much linear response as possible. While this simplifies the wavefront reconstruction and control, one can extend the design space of such wavefront sensors by allowing for nonlinearities \cite{2016OExpr_haffert_godwfs}. The response of the PWFS is generally already nonlinear for large aberrations and/or in the presence of residual turbulence \cite{1999A&A...350L..23R_ragazzoni_pwfs_sensitivity,2001A&A...369L...9E_esposito_pwfs_residuals}. This problem can be alleviated by algorithms that track and correct for the effective gain \cite{2019A&A_deo_optical_gain_modal, 2020A&A...644A...6C_chambouleyron_pwfs_optical_gains, 2021A&A_deo_correlation_locking_filter, 2021JATIS_haffert_ddspc} or through nonlinear reconstruction algorithms\cite{2018ApOpt_hutterer_nonlinear_pyramid,2018JOSAA_frazin_pyramid_nonlinear,2020JATIS_shatokhina_pwfs_reconstruction_review}. A promising area of research for mitigating these nonlinearities is the use of neural networks for learning a nonlinear mapping between wavefront sensor measurements and wavefront \cite{ 2018SPIE10703E..1FS_swanson_cnn_pred, 2020OExpr..2816644L_landman_nlwfs,2020NatCo_norris_photonicwfs, 2021MNRAS.505.5702O_orban_ml_fpwfs}, or for nonlinear control \cite{ 2021JATIS_landman_rl,2021JATIS_wong_nonlinear_predictivecontrol, 2022OExpr_pou_rl, 2022arXiv220507554N_nousiainen_rlcontrol}. Furthermore, the similarities between optical systems and Neural Networks have lead to studies exploiting automatic differentiation algorithms, initially developed for training NNs, for optimizing elements in the optical system \cite{2021ApJ_pope_autodiff, 2021JOSAB..38.2465W_wong_morphine} or more efficient wavefront control \cite{2021JATIS_Will_jacobianfree_control,2021SPIE11823E..0VW_scott_algodiff_hicat}. Automatic differentiation allows us to obtain gradients with respect to the free design parameters, even for complex optical systems with multiple elements and planes.

In this work, we attempt to numerically optimize the focal plane mask in a Fourier-filtering wavefront sensor using automatic differentiation. First, we optimize the sensitivity of the WFS for different apertures and wavefront distributions. After that, we will directly minimize the residual wavefront after correction, by jointly optimizing the WFS and reconstructor. These optimized wavefront sensors (OWFS) are obtained by connecting a differentiable model of the optical system with a differentiable model of the reconstruction algorithm. We then alternatingly improve them by performing a gradient step with a gradient-based optimizer. With this approach, one can also use a nonlinear reconstructor (e.g. a Neural Network), thereby allowing for nonlinearities in the response of the WFS.  While we apply it here to a Fourier-filtering wavefront sensor architecture, our method can generally be used to optimize the parameters of any kind of wavefront sensor system with free parameters.

\section{METHODS}
\subsection{DIFFERENTIABLE OPTICAL SIMULATION}
We simulate our optical system using a custom framework based on HCIPy \cite{2018SPIE10703E_por_hcipy}. All operations are implemented using TensorFlow \cite{2016arXiv160304467A_tensorflow}, such that we can retrieve the gradients with respect to our free parameters using automatic differentiation. The optical setup of a Fourier Filtering wavefront sensor consists of a set of lenses with a focal plane mask inbetween\cite{2016Optic...3.1440F_fauvarque_fourier_formalism}. The resulting image $I$ of the wavefront sensor for a specific aperture $A$, input phase $\phi$ and focal plane mask $m$, is then given by:
\begin{equation}
    I(\phi; m) = |\mathcal{F}^{-1}[m \times \mathcal{F}(A \exp{i\phi})]|^2,\label{eq:ff_wfs}
\end{equation}
where $\mathcal{F}$ is the Fourier Transform operator. The propagation was implemented using a Matrix Fourier Transform \cite{2007OExpr..1515935S_soummer_fast_propagation} in TensorFlow, and thus consist of a single large matrix multiplication. Eq. \ref{eq:ff_wfs} shows that the relation between incoming phase and WFS image is indeed generally nonlinear. While the current framework allows for polychromatic simulations, we will only consider monochromatic light at 1 micron in this work. For all simulations, we assume we have total of $10^6$ photons in each frame and add photon noise accordingly.

\subsection{WAVEFRONT RECONSTRUCTION}
\begin{figure}[htbp]
    \centering
    \includegraphics[width=\linewidth]{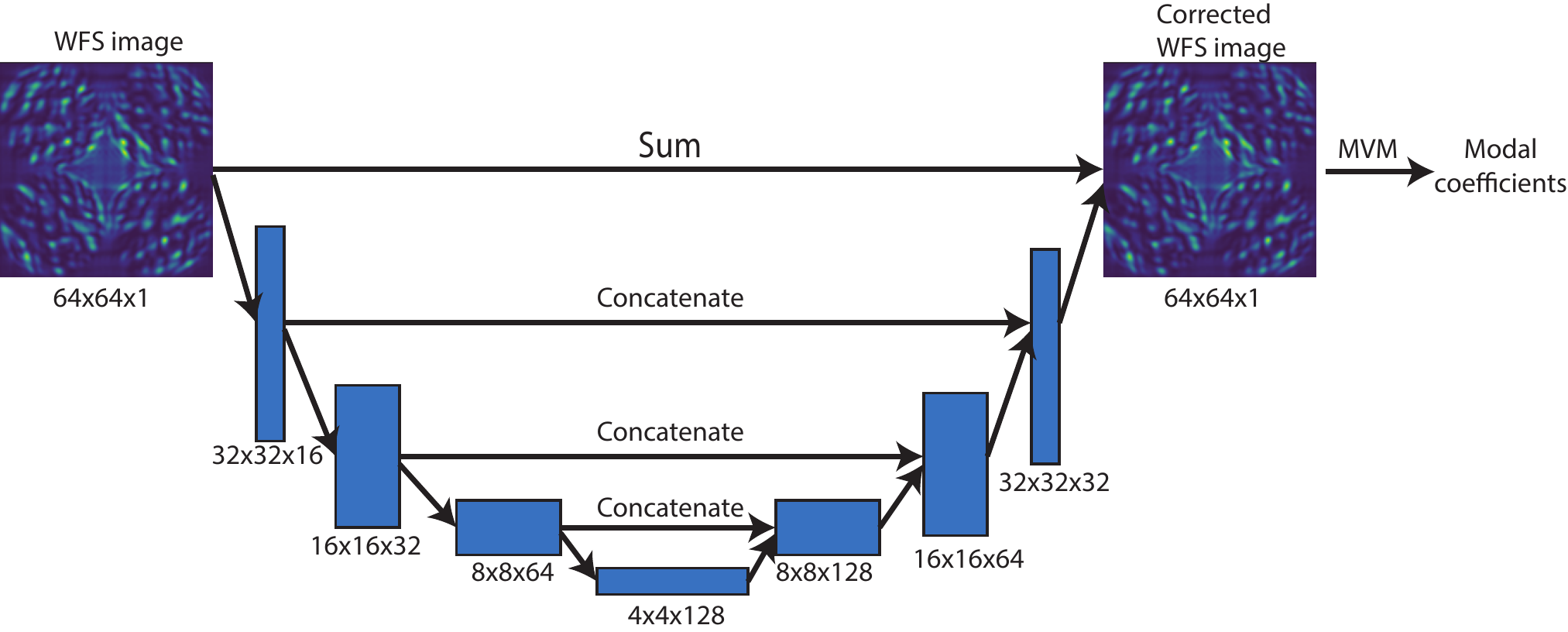}
    \caption{Architecture of the Convolutional Neural Network used for the nonlinear reconstruction. All layers use a 3x3 kernel size and leaky ReLU activation, except for the output image, which has a linear activation.}
    \label{fig:cnn_architecture}
\end{figure}

The inverse operation of retrieving the wavefront from the wavefront measurements can be written as:
\begin{equation}
    \phi_{rec}(I) = R(I;\theta),
\end{equation}
where $R$ is the reconstruction model with free parameters $\theta$. This reconstruction can be done in a modal basis of choice. One can find the best set of parameters $\theta$ for the reconstructor through calibration. The most common approach is to have a linear reconstruction model:
\begin{equation}
    R(I,\theta) = \theta I,
\end{equation}
for which $\theta$ can be found by through a regularized linear least-squares regression. For example, with Tikhonov regularization, the linear least squares solution is given by:
\begin{equation}\label{eq:least_squares}
     \theta = (A^T A + \lambda \mathbb{I})^{-1}A^T, 
\end{equation}
where $A$ is the measured interaction matrix in the chosen modal basis, $\lambda$ a regularization parameter and $\mathbb{I}$ the identity matrix. However, we are not limited to linear models and are generally free to choose the parameterization of $R$. For example, one can use nonlinear function approximators, such as  Neural Networks (NN), for this. In this case, the optimal parameters of the reconstructor can no longer be found using a linear least squares. Instead, gradient-based optimizers are commonly used to find good sets of parameters for these models. Here, we will use a Convolutional Neural Network (CNN) as our nonlinear reconstructor. These CNN's have been used in many inverse problems involving images, and have led to impressive results\cite{2015arXiv151203385H_resnet, 2015arXiv150504597R_unet}. The CNN architecture used here is based on U-net \cite{2015arXiv150504597R_unet}, and is illustrated in Fig. \ref{fig:cnn_architecture}. The output of our CNN has the same shape as the original image and consists of a correction term on the intensity image of the WFS. This correction term is added to the intensity image,  which is then propagated through a MVM to reconstruct the wavefront in the chosen basis. This approach was motivated by the hybrid approach in Ref. \citenum{2020OExpr..2816644L_landman_nlwfs}. This also decouples the nonlinearity correction of the WFS and the projection onto a modal basis, allowing flexibility in the choice of modal basis after training the CNN. 

\subsection{OPTIMIZATION}
Using our differentiable optical simulations and reconstruction, we can optimize the free parameters of our wavefront sensor and reconstructor using gradient-based optimization methods. In particular, we want to optimize the Fourier filtering focal plane mask for 1) maximum sensitivity and 2) minimum residual wavefront error.

\subsubsection{Maximum sensitivity}\label{sec:optimization_sensitivity}
First, we will directly optimize for the sensitivity of the wavefront sensor.  We use the definition from Chambouleyron et al. 2022\cite{2021A&A...650L...8C_chambouleyron_zernike2} for the sensitivity $s$ of the wavefront sensor with respect to a wavefront mode $\phi_i$:
\begin{equation}\label{eq:sensitivity}
    s(\phi_i) = \frac{||\delta I(\phi_i;m) ||_2}{||\phi_i||_2},
\end{equation}
where $\delta I(\phi_i, m)$ is obtained using a "push-pull" of the wavefront mode $\phi_i$ for a small amplitude $\epsilon$:
\begin{equation}
    \delta I(\phi_i;m) = \frac{I(\epsilon\phi_i; m)- I(-\epsilon\phi_i;m)}{2\epsilon}.
\end{equation}
Chambouleyron et al. 2022\cite{2021A&A...650L...8C_chambouleyron_zernike2} showed that one can increase the sensitivity to higher spatial frequencies by increasing the size of the Zernike dot, at the cost of decreased sensitivity to tip-tilt. The wavefront sensor with the optimal effective sensitivity thus depends on the modes we want to be sensitive to. Furthermore, the sensitivity of one mode is not independent of other modes, as there is nonlinear cross-talk. This is for example shown by the reduction in sensitivity of the PWFS under residual turbulence \cite{2019A&A_deo_optical_gain_modal}. To account for both effects, we generate different phase screen realizations for a given statistical distribution and optimize for the expected sensitivity over this distribution:
\begin{equation}
    \textrm{Loss}(m) = -\mathbb{E}_{\phi_{in}}\left[s(\phi_{in}) \right].
\end{equation}
Here, $\mathbb{E}$ denotes the expectation value over the chosen wavefront distribution. The expectation value is taken over batches of 32 randomly generated wavefronts from this distribution. One issue is that this metric only considers the response, and does not consider our ability to reconstruct the wavefront from the measurements. This could for example lead to WFS's with high cross-talk, a very nonlinear response, or a small dynamic range. 

\subsubsection{Minimum residual phase through joint optimization}\label{sec:joint_optimization}
To solve this, we will also jointly optimize the WFS and reconstructor.  Using this, we can directly optimize for the residual error in the wavefront estimation.  This way we take into account all relevant effects, including cross-talk, dynamic range and noise propagation through the reconstruction. For maximum Strehl ratio, we have to minimize the residual mean squared error:
\begin{equation}
    \textrm{Loss}(m,\theta) = \mathbb{E}_{\phi_{in}}\left[ \langle (\phi_{in} - \phi_{rec})^2 \rangle \right],
\end{equation}
with
\begin{equation}
    \phi_{rec} = R(I(\phi_{in}; m); \theta).
\end{equation}
Here, $\langle \cdot \rangle$ denotes the average over the aperture. It is best to take the expectation value over the expected residual phase aberrations that are seen in closed-loop on-sky. However, this distribution is hard to estimate and inherently depends on the properties of the WFS and reconstructor itself.  Instead, we use the loss function from Landman et al. 2021\cite{2020OExpr..2816644L_landman_nlwfs}, which uses the following relative loss:
\begin{equation}
    \textrm{Loss}(m,\theta) = \mathbb{E}\left[ \frac{\langle (\phi_{in} - \phi_{rec})^2 \rangle }{\langle\phi_{in}^2\rangle+\epsilon}\right],
\end{equation}
where $\epsilon$ is a term to avoid diverging loss for small input wavefront aberrations.  This loss function makes sure that we get good performance both in the large and small aberration regime, and that the system converges in closed-loop.

\begin{figure}[htbp]
    \centering
    \includegraphics[width=0.9\linewidth]{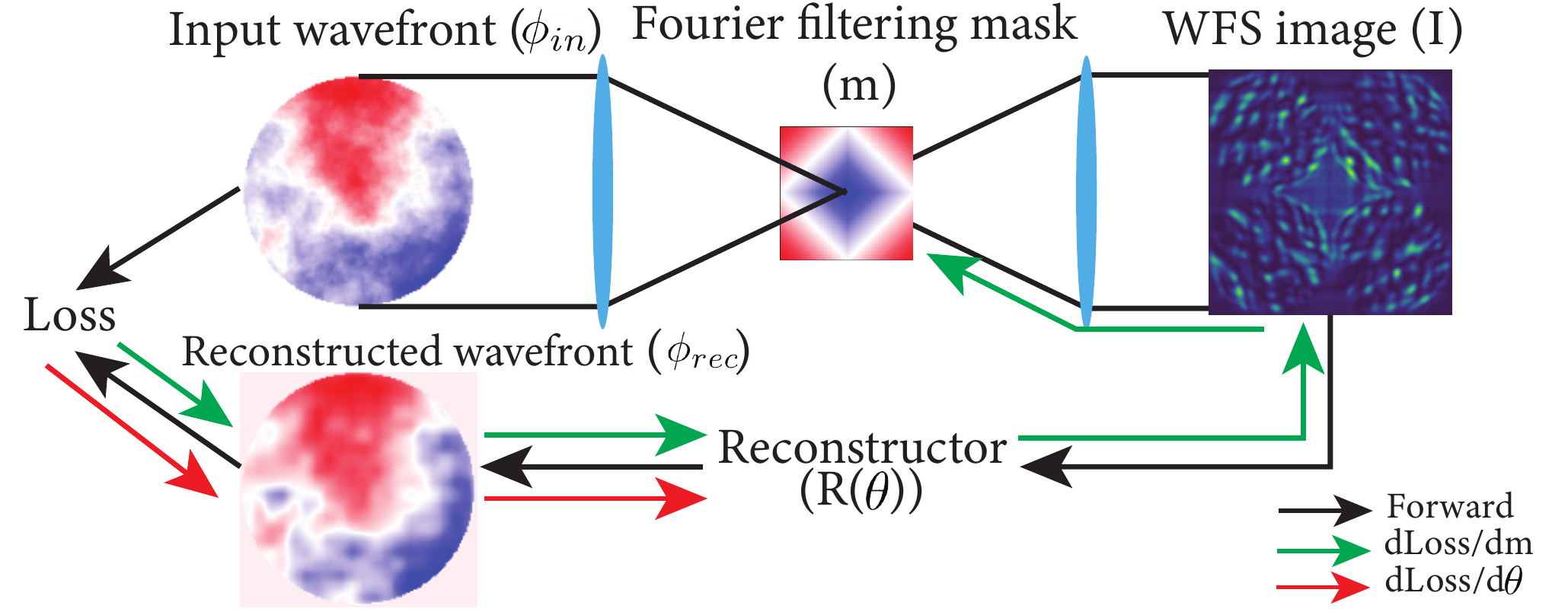}
    \caption{Schematic overview of the forward model of the Fourier-filtering wavefront sensor and how the gradients are backpropagated for the joint optimization.}
    \label{fig:ff_wfs_grad}
\end{figure}

Since both our forward model $I(\phi_{in}; m)$ and reconstructor $R(I;\theta)$ are differentiable, we can then find the gradients with respect to $\theta$ and $m$ using automatic differentiation. This is illustrated in Fig. \ref{fig:ff_wfs_grad}.  When the gradients are known, we can optimize the mask and reconstructor using a gradient-based optimizer.  We alternatingly do a gradient step for the mask and then three steps for reconstructor to make sure that the reconstructor is properly updated for the new mask. The optimization is done using the Adam algorithm \cite{2014arXiv1412.6980K_adam} with a learning rate of 0.01 and 0.001 for the focal mask phase and reconstructor respectively, and again a batch size of 32. Every 500 gradient updates we decay both learning rates with a factor 0.96. 

\label{sec:intro}  
\section{RESULTS}
\subsection{MAXIMUM SENSIVITY}\label{sec:results_sensitivity}
\begin{figure}[htbp]
    \centering
    \includegraphics[width=\linewidth]{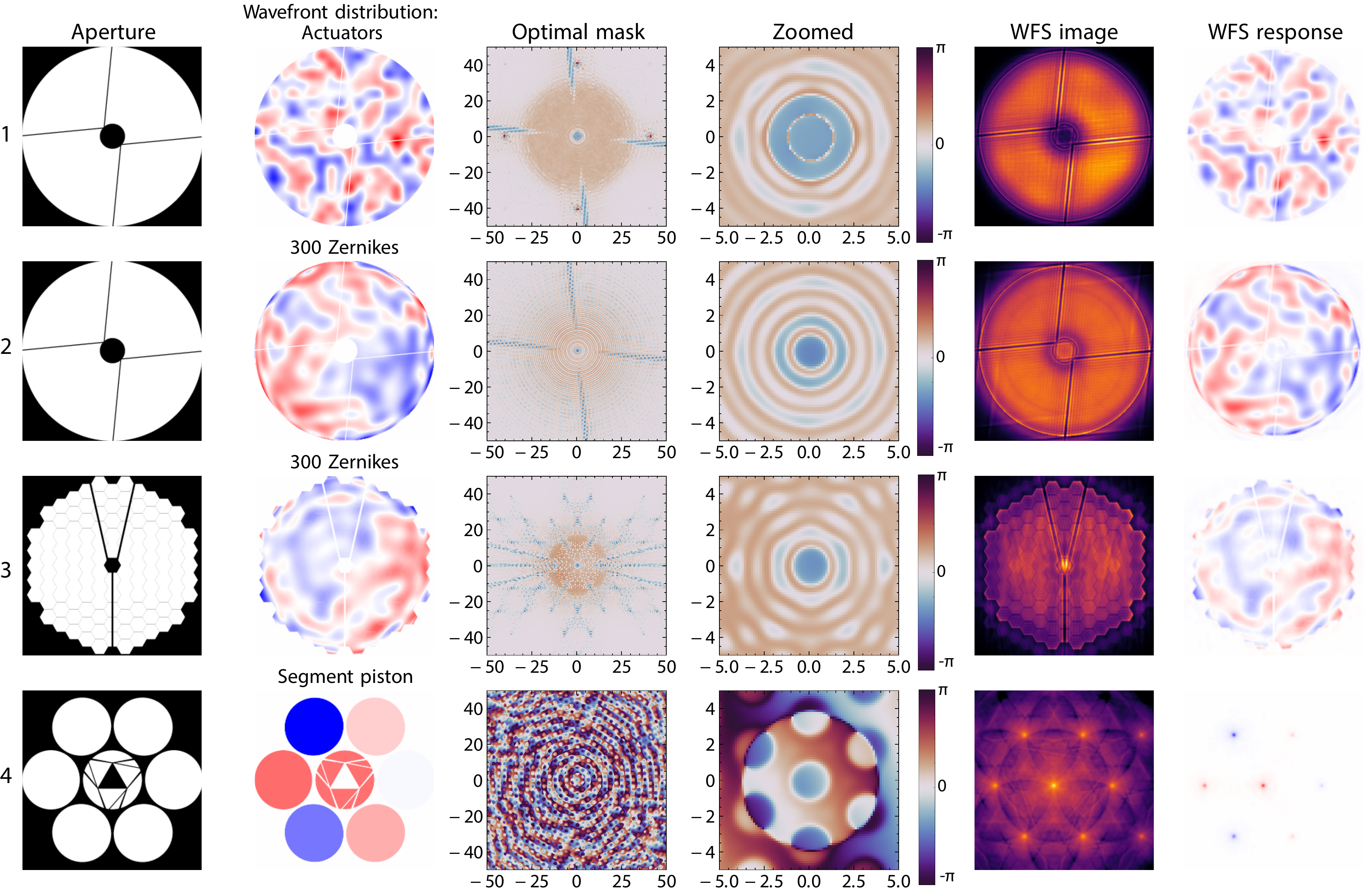}
    \caption{From left to right: The used aperture, a sample from the wavefront distribution, the optimal mask, the optimal mask zoomed around the central Airy core, the wavefront sensor image, and the wavefront sensor response for the sampled wavefront aberration. This is shown for each of the four test cases specified in section \ref{sec:results_sensitivity} from top to bottom.}
    \label{fig:opt_sensitivity}
\end{figure}

First, we optimize for maximum sensitivity to a certain wavefront distribution, as explained in section \ref{sec:optimization_sensitivity}. We have tested four different cases: 
\begin{enumerate}
    \item The VLT aperture with maximum sensitivity w.r.t. the actuators modes, for which we have used a 40x40 DM similar to current high-contrast imaging instruments. This represents the case with a flat spatial power spectrum within the control radius.
    \item The VLT aperture with maximum sensitivity w.r.t the first 300 Zernikes distributed according to Ref. \citenum{1976JOSA_noll_zernike}.
    \item Same as 2 but with the LUVOIR-A aperture.
    \item The GMT pupil with optimized sensitivity w.r.t. segment piston.
\end{enumerate}
 We optimize for very small abberations with an RMS of 0.1 nm. The resulting optimal masks with corresponding WFS image and response are shown in Fig. \ref{fig:opt_sensitivity}. This shows that the optimization ends up on masks similar to the ZWFS, with a $\pi/2$ phase shift for the Airy core. When optimizing for the actuator modes (flat power spectrum), we get a larger dot diameter than when optimizing for the Zernike modes. This is in agreement with Chambouleyron et al. 2021 \cite{2021A&A...650L...8C_chambouleyron_zernike2}, who find that increasing the dot diameter increases the sensitivity to higher spatial frequencies. Furthermore, the diffraction structures due to the spiders also get a $\pi/2$ phase shift, and light inside the control radius gets a different phase shift than outside. For the optimized mask for the Zernike modes, we get a copy of the point spread function (PSF), with a slowly declining phase amplitude going to the outer Airy rings. If we change the aperture shape, e.g. in the LUVOIR-A case, the optimal masks changes according to the new structure of the PSF. Finally, the case for optimal sensitivity to differential piston for the GMT results in a more exotic phase mask in the focal plane. This case is just to illustrate that the methodology can in principle be adjusted to different science cases.
 
\begin{figure}[htbp]
    \centering
    \includegraphics[width=\linewidth]{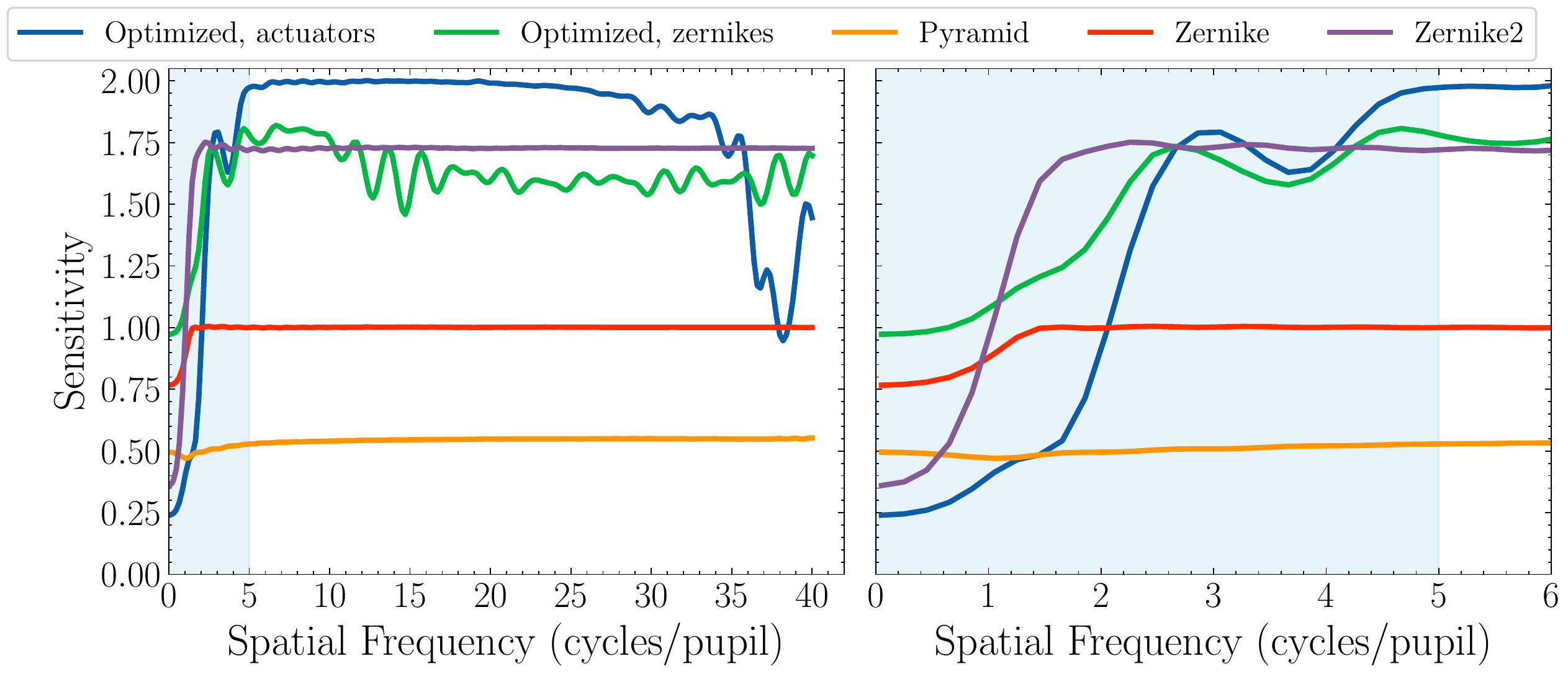}
    \caption{Sensitivity of the different wavefront sensors to aberrations of different spatial frequencies. The Pyramid is the unmodulated case here.}
    \label{fig:sensitivity_curve_opt}
\end{figure}

We show the sensitivity curve of the two optimized wavefront sensors (OWFS) for the VLT aperture in Fig. \ref{fig:sensitivity_curve_opt}, following the definition from Eq. \ref{eq:sensitivity}. We also show the sensitivity curves of the unmodulated PWFS, the classical ZWFS with a 1.06 $\lambda/D$ dot \cite{2013A&A...555A..94N_ndiaye_zelda}, and the Zernike2 WFS (Z2WFS) with a 2 $\lambda/D$ dot size\cite{2021A&A...650L...8C_chambouleyron_zernike2}. We again see the enhanced sensitivity of the Z2WFS over the ZWFS for spatial frequencies $>1$ cycle/pupil, at the cost of decreased sensitivity to tip/tilt. The OWFS for the actuator modes follows this trend, with very low sensitivity to small spatial frequencies, but a maximum sensitivity of 2 to mid spatial frequencies, after which it then again drops of. The OWFS for the Zernike modes retains sensitivity to tip-tilt, while also providing increased sensitivity to the higher spatial frequencies over the classical Zernike. It appears that this OWFS has a lower sensitivity than the Z2WFS for higher spatial frequencies, but this may simply be the result of our choice of only including the first 300 Zernike modes in the optimization. A large caveat to these results is that these are monochromatic simulations and the improvement may degrade when working over a larger bandwidth. While the chromaticity of the phase mask may be avoided using liquid crystal technology \cite{2017SPIE10400E_doelman_liquid_crystal, 2019OptL_doelman_vectorzernike}, the phase mask would not exactly align with the PSF shape anymore, which could lead to decreased sensitivity.

\begin{figure}[htbp]
    \centering
    \includegraphics[width=\linewidth]{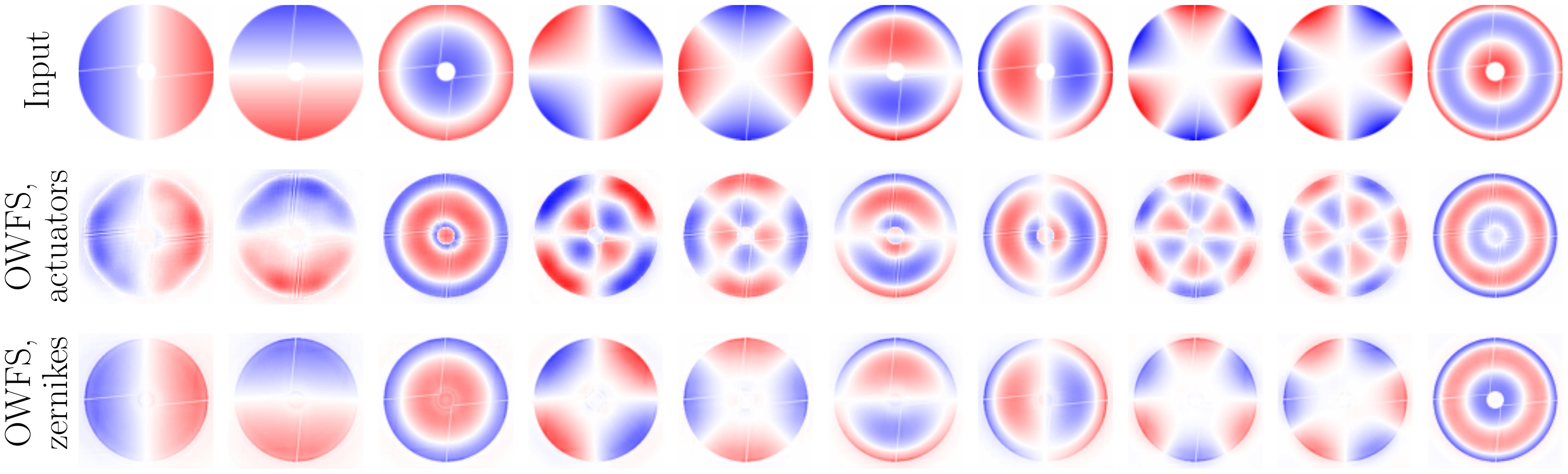}
    \caption{Response of the optimized wavefront sensors for the VLT aperture to the first 10 Zernike modes.}
    \label{fig:response}
\end{figure}

\begin{figure}[htbp]
    \centering
    \includegraphics[width=0.5\linewidth]{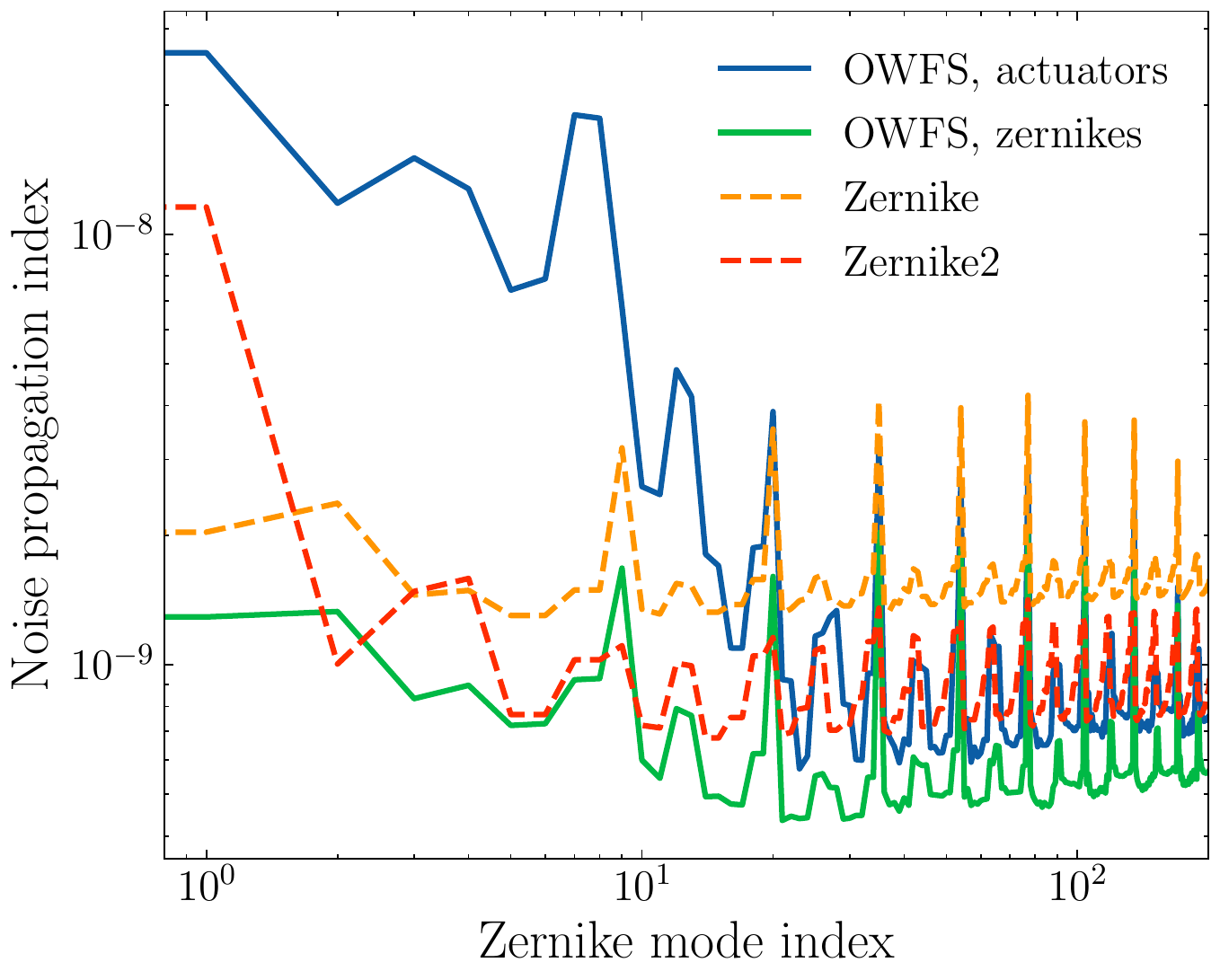}
    \caption{Diagonal elements of $(A^T A)^{-1}$ for the different wavefront sensors for the first 300 Zernike modes, showing the noise propagation.}
    \label{fig:noise_propagation}
\end{figure}

Additionally, the sensitivity as defined in Eq. \ref{eq:sensitivity} is not the entire story, as we need to be able to reconstruct the wavefront from the measurements. If the response of different modes is very correlated, the reconstruction might lead to a large noise amplification. The response to the first 10 Zernike modes of the OWFS optimized for the VLT aperture is shown in Fig. \ref{fig:response}. This shows that the OWFS for the actuator modes sees lower order modes (e.g. astigmatism) as higher order modes. This may lead to degeneracies between different modes and thus noise amplification with the reconstruction. To investigate this, we use the approach from Fauvaurque et al. 2015 \cite{2015OptL...40.3528F_fauvarque_flattened_pwfs}, and look at the diagonal entries of $(A^T A)^{-1}$ for the first 300 Zernike modes, with $A$ the interaction matrix. This is shown in Fig. \ref{fig:noise_propagation}. This shows that while the OWFS for the actuators gives the highest sensitivity according to Eq. \ref{eq:sensitivity}, it has higher noise propagation than the one obtained for the Zernike modes. The OWFS for the Zernike modes has lower noise propagation than the classical ZWFS for all modes, even for tip-tilt, and thus seems to be a more sensitive WFS than the ZWFS and Z2WFS. Still, the sensitivity defined as in Eq. \ref{eq:sensitivity} is not the correct metric to optimize for.

\subsection{JOINT OPTIMIZATION}
\begin{figure}[htbp]
    \centering
    \includegraphics[width=\linewidth]{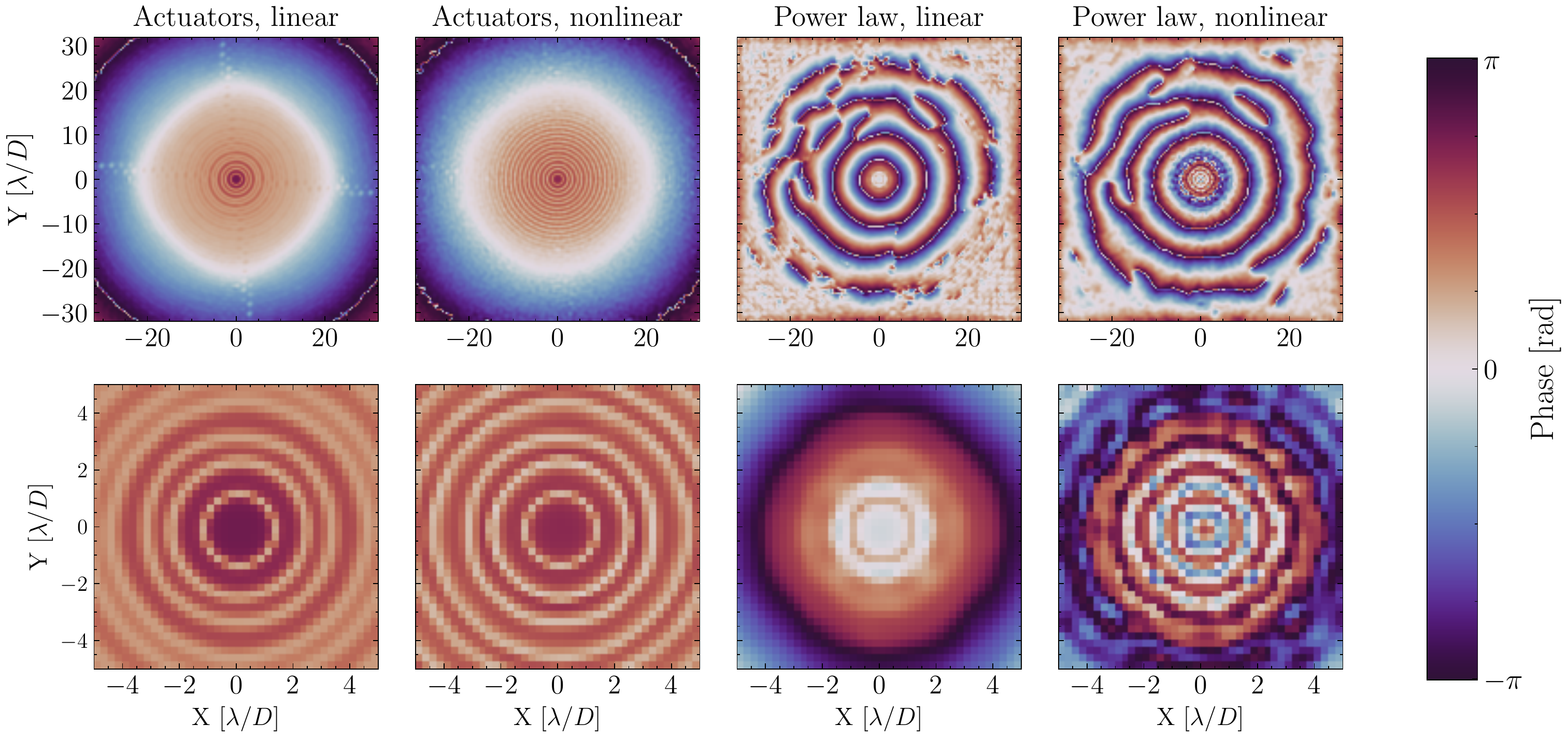}
    \caption{Resulting focal mask phase after jointly optimizing the mask and reconstructor for two different input distributions: 1. Random combination of actuator pokes, 2. Kolmogorov turbulence.}
    \label{fig:opt_mask_joint}
\end{figure}

\begin{figure}[htbp]
    \centering
    \includegraphics[width=\linewidth]{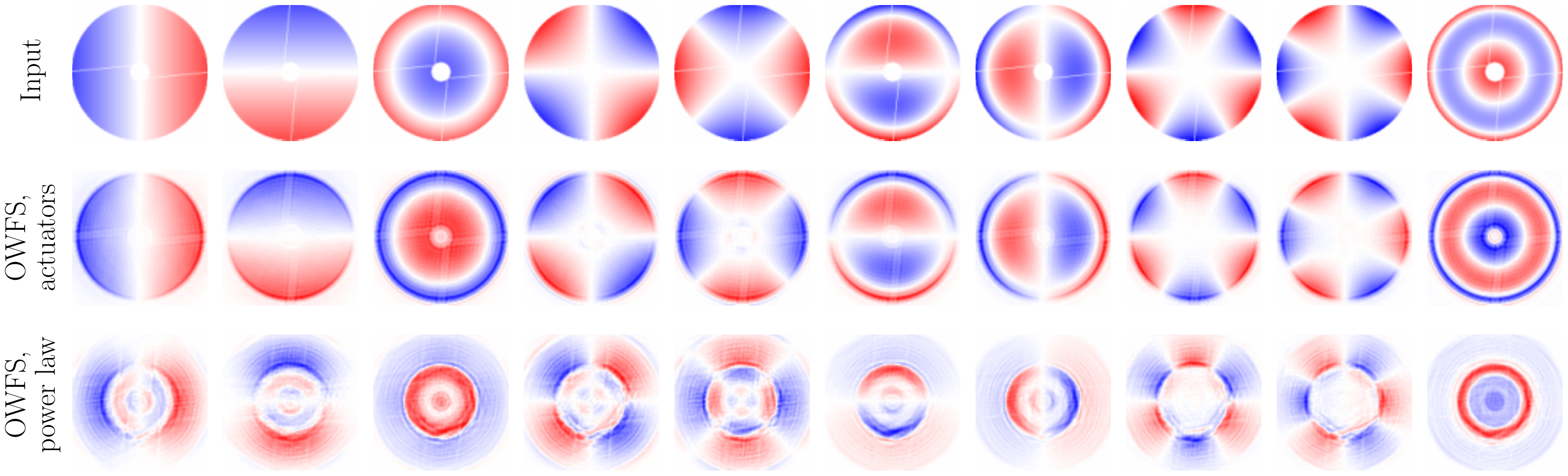}
    \caption{Response of the jointly optimized wavefront sensors to the first 10 Zernike modes.}
    \label{fig:joint_responses}
\end{figure}

Therefore, we will here optimize directly for the residual wavefront error by considering the wavefront reconstruction, as described in section \ref{sec:joint_optimization}. We do this for both a linear reconstructor and the CNN shown in Fig. \ref{fig:cnn_architecture}.  We will consider two edge cases for the input distribution: The case of Kolmogorov turbulence with a -11/3 power spectral density (PSD), and a random combination of actuator pokes, which is a flat PSD within the control radius. In reaility, the closed-loop distribution will be somewhere in between those cases. The RMS of the input wavefront is sampled from a log-uniform distribution between 1 nm and 3 $\mu$m, such that we optimize over the full range of aberration scales that are seen in closed-loop operation. We only consider the VLT aperture here and the resulting optimized focal plane masks are shown in Fig. \ref{fig:opt_mask_joint}. For the actuator modes, we get a very similar result as for the maximum sensitivity, with a $\pi/2$ phase shift for the PSF. There is an additional defocus term, which might help with the linearity range of the sensor. The result for the nonlinear optimization is almost the same as for the linear one. For the power-law turbulence in the linear case, we end up with a phase mask that is a combination of an axicone and a phase shift of the Airy core and first Airy ring, similar to the ZWFS. This combination of a pyramidal-like structure combined with a ZWFS is similar to the bright/dark PWFS concept from Gerard et al. 2022 \cite{2021SPIE11823E..1BG_gerard_bright_pwfs}. The response of the optimized masks for the linear reconstructors is shown in Fig. \ref{fig:joint_responses}.

\begin{figure}
    \centering
    \includegraphics[width=0.49\linewidth]{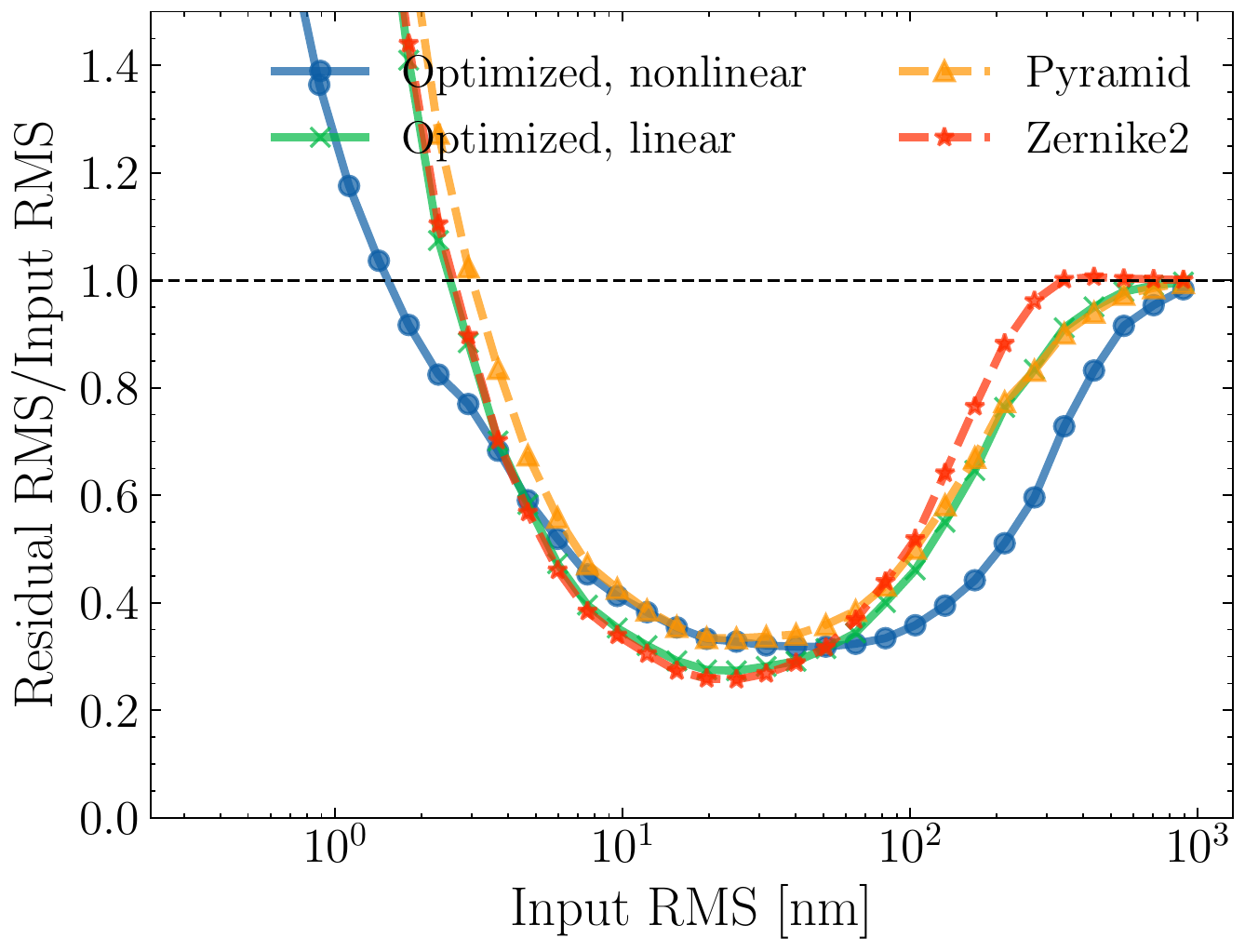}
    \includegraphics[width=0.49\linewidth]{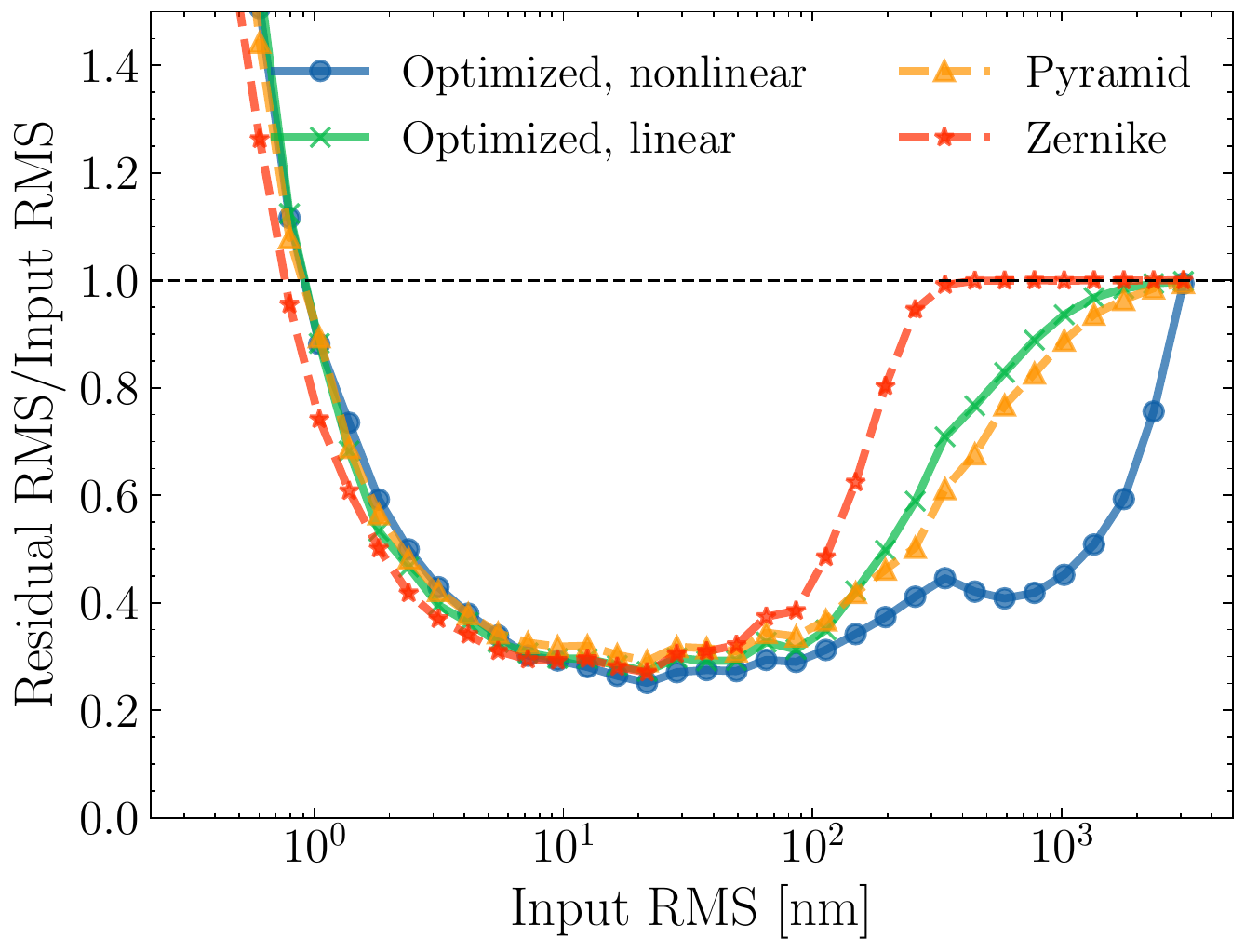}
    \caption{Comparison of the performance of the different wavefront sensors. It shows the "improvement factor", which is residual RMS divided by the input RMS, as a function of input RMS. Left: Input distribution consists of a random combination of actuator pokes. Right: Input distribution consists of Kolmogorov turbulence.}
    \label{fig:gain_curve}
\end{figure}

Finally, we study the performance of the various masks and reconstructors by looking at the residual wavefront error. Fig. \ref{fig:gain_curve} shows the residual RMS divided by the input RMS as a function of the input RMS. This factor shows the improvement in wavefront quality we get after a single correction. If this ratio reaches 1, we can not improve the wavefront and have reached the limits of the wavefront sensor. The limit for small RMS is set by the noise propagation, while the limit for large RMS is set by the capture range. These curves show that the optimization can lead to wavefront sensors with enhanced capture range, especially when using a nonlinear reconstructor, with minimal loss in sensitivity. Future work will compare the performance of these optimized wavefront sensors in more detail.

\section{CONCLUSIONS}
We have shown that optimizing the focal plane mask in a Fourier-filtering wavefront sensors can lead to more sensitive and better performing wavefront sensors. We find that the mask for optimal sensitivity depends on the modes one wants to be sensitive to and the shape of the point spread function. Our optimization leads to wavefront sensors that are more sensitive than the Zernike and Pyramid wavefront sensors and have a sensitivity close to the theoretical optimum. However, we find that sensitivity is not the correct metric to optimize for, as it does not consider noise propagation through the reconstruction. By jointly optimizing the wavefront sensor and reconstructor, one considers all relevant effects and directly minimizes the residual wavefront aberrations. Furthermore, by using a nonlinear reconstructor, we can extend the design space of these WFS's and extend the capture range. Initial results from joint optimization show that this can lead to wavefront sensors with improved capture range. Future work will explore polychromatic simulations, optimal modulation schemes, and closed-loop performance simulations.

\section*{ACKNOWLEDGEMENTS}
R.L. acknowledges funding from the European Research Council (ERC) under the European Union's Horizon 2020 research and innovation program under grant agreement No 694513. E.H.P. is supported by the NASA Hubble Fellowship grant \#HST-HF2-51467.001-A awarded by the Space Telescope Science Institute, which is operated by the Association of Universities for Research in Astronomy, Incorporated, under NASA contract NAS5-26555. S.Y.H. is supported by NASA through the NASA Hubble Fellowship grant \#HST-HF2-51436.001-A awarded by the Space Telescope Science Institute, which is operated by the Association of Universities for Research in Astronomy, Incorporated, under NASA contract NAS5-26555.

\bibliography{report} 
\bibliographystyle{spiebib} 

\end{document}